\documentclass{article}
\usepackage{epsfig}

\date{\today}

\usepackage{amsmath}
\usepackage{amssymb}
\usepackage[hang,nooneline,scriptsize]{subfigure}

\makeatletter
\let\@fnsymbol\@arabic
\makeatother
\begin{document}
\title{From topological to non-topological solitons:\\
kinks, domain walls and $Q$-balls in a 
scalar field model\\ with non-trivial vacuum manifold}

\author{
{\large Yves Brihaye}$^{\dagger}$\footnote{email: yves.brihaye@umons.ac.be} , 
{\large Adolfo Cisterna }$^{\ddagger}$\footnote{email:adolfo.cisterna.r@mail.pucv.cl} ,
{\large Betti Hartmann }$^{\diamond}$\footnote{email:bhartmann@ifsc.usp.br},
{\large Gabriel Luchini}$^{\star}$\footnote{email:gabriel.luchini@gmail.com}
\\ \\
$^{\dagger}${\small Physique-Math\'ematique, Universit\'e de
Mons, 7000 Mons, Belgium}\\ 
$^{\ddagger}${ \small Instituto de Ciencias F\'{\i}sicas y Matem\'{a}ticas, Universidad
Austral de Chile, Casilla 567, Valdivia, Chile}  \\
$^{\diamond}${ \small Instituto de F\'{\i}sica de Sao Carlos (IFSC), Universidade de S\~ao Paulo (USP), 
CEP 13560-970 567, S\~ao Carlos (SP), Brazil} \\
$^{\star}${ \small Departamento de F\'isica, Universidade Federal do Esp\'irito Santo (UFES), CEP 29075-900, Vit\'oria-ES, Brazil} 
}


\date{\today}

 \maketitle
\begin{abstract} 
We consider a scalar field model with a self-interaction potential that possesses
a discrete vacuum manifold. We point out that this model allows for both topological as well
as non-topological solitons. In $(1+1)$ dimensions both type of solutions have finite energy,
while in $(3+1)$ dimensions, the topological solitons have finite energy per unit area only and correspond to domain walls. 
Non-topological solitons with finite energy do exist in $(3+1)$ dimensions 
due to a non-trivial phase of the scalar field and an associated $U(1)$ symmetry of the model, though. We construct
these so-called $Q$-ball solutions numerically, point out the differences to previous studies
with different scalar field potentials and also discuss the influence of a minimal coupling to both gravity as well as a $U(1)$ gauge field. In this
latter case, the conserved Noether charge $Q$ can be interpreted as the electric charge of the solution. 
\end{abstract}

\section{Introduction}
Attributing a scalar quantity to each space-time point, scalar fields are one of the simplest fields possible. Often used in effective
descriptions or serving as simpler toy models of physical phenomena, they, however, arise as well as 
fundamental constituents of many theoretical physics models. The prime example is String Theory,
which contains scalar fields as fundamental pieces of its scaffolding. Indeed, in many models beyond the Standard Model of
Particle Physics,
a scalar field, the dilaton, appears as an irreducible representation of the first excited states \cite{Polchinski:1998rq, Tong:2009np}. 
Scalar fields also arise naturally in other theories like in Kaluza-Klein theory, where gravity and electrodynamics can 
be formulated as different manifestations of the gravitational field in a five-dimensional space-time. 
After dimensional reduction to four space-time dimensions, a scalar field appears \cite{Fon9}. Scalar fields can 
also be found in Supergravity theories as auxiliary fields ensuring off-shell realizations of supersymmetry \cite{Freedman:2012zz} 
and in several cosmological models in order to drive inflation or being part of dark matter models \cite{Joyce:2014kja}. 
A very appealing case is the Starobinsky model which is conformally equivalent to gravity with a minimally coupled scalar 
field \cite{Starobinsky:1980te}. In this model, the massive scalar state, usually called scalaron, possesses 
a very particular potential able to drive the inflationary process of the early universe, describing a 
slow-roll transition from a de-Sitter phase to a Minkowski flat phase.
Due to the wide variety of applications and due to the discovery 
of a fundamental scalar particle in nature \cite{atlas}, scalar fields have been studied increasingly in the past years. 
\\
In classical theories (as opposed to quantum theories) for scalar fields, soliton configurations are of particular relevance. A soliton refers to a non-singular solution of a non-linear equation which, after collecting all its features, can be pictured as a travelling wave of (almost) unchangeable shape, whose energy remains finite in a finite region of space, so that it resembles a localised lump.
The interesting feature concerning these solutions is that even as classical waves, they behave in many aspects like particles, especially 
when their scattering is considered \cite{Manton:2004tk}.
\\
The concept of solitons was introduced in the context of a scalar field theory in $(1+1)$--dimensional Minkowski space-time \cite{zabusky}. 
The stability of the solutions in this particular model is explained by the high number of constraints (in fact, there are infinitely many of them) 
that this solution has to satisfy. These constraints appear as conserved charges - in principle, 
completely unrelated to Noether charges - and, being infinitely many, they define the theory as integrable. Such an integrability concept, however, does not go beyond one spatial dimension, not to mention the difficulties in discussing it in curved space(-time). 
\\
On the other hand, in higher dimensions there are soliton-like configuration, i.e., solutions of the field equations in the form of localised lumps. It is the integrable structure of certain non-linear field theories that enables the development of methods for construction of soliton solutions \cite{laf_wojtek}. Since such structure is not present for theories in higher dimensional space-time, there are no well established analytical methods to get the solitonic solutions there, and therefore most of these solutions are obtained numerically.
\\
For spatial dimension larger than one topological solitons possess an invariant that classifies the mapping between 
two manifolds $\Psi: \mathcal{X} \longmapsto\mathcal{Y}$, $\mathcal{X}$ being the manifold of the degrees of freedom of the solution 
with local coordinates $\vec{x}$ and $\mathcal{Y}$ the physical $D$--dimensional space with local coordinates $\vec{y}$, 
and $\textrm{dim}\mathcal{X} = \textrm{dim}\mathcal{Y}$. This topological invariant, also referred to as topological charge, 
is the pull-back in $\mathcal{X}$ of a volume form $\Omega = \beta(\vec{y})dy^1 \wedge \dots \wedge dy^{D}$ in $\mathcal{Y}$ normalised to unity and reads:
\begin{equation}
N = \int_{\mathcal{X}}\beta(\vec{y}(\vec{x}))\Big\vert{\frac{\partial \vec{y}}{\partial \vec{x}}}\Big\vert dx^{1}\wedge \dots \wedge dx^{D}.
\end{equation}
\\
Thus, the time evolution of a configuration is viewed as an homotopy between its initial and final state and the solitons are characterised in an equivalence class of maps $\Psi$.
\\
The possibility of existence of pure scalar field solitons in more than one spatial dimension is ruled out by Derrick's theorem (see e.g. \cite{Manton:2004tk}). So, in higher dimensions, a theory that exhibits solitons involves other fields (e.g. gauge fields) as well. Another possibility is that the scalar field has internal degrees of freedom, e.g. a non-trivial phase.
\\
Indeed, this last option gives rise to so-called {\it $Q$-balls} \cite{Coleman:1985ki}.
These configurations are soliton solutions of a complex scalar field theory which interacts through a suitable self-interaction potential. They
are non-topological in nature \cite{Lee:1991ax} and it is 
the invariance of the action under global phase transformations of the scalar field which endorses the theory with a conserved Noether charge 
$Q$ interpreted as the particle number. Once the theory is gauged, this Noether charge multiplied by the coupling constant of the gauge field becomes the total charge of the soliton.
Rotating and non-rotating $Q$-balls have been shown to exist with polynomial self-interaction containing up to 6th order 
terms \cite{Volkov:2002aj,Kleihaus:2005me} as well as for an exponential self-interaction potential motivated by supersymmetric 
extensions of the standard model \cite{Kusenko:1997zq,Kusenko:1997ad,Hartmann:2012gw}.
\\
Gravitating solitons in this context, namely {\it boson stars}, do not require self-interaction of the scalar field. 
Indeed boson stars can be constructed with a scalar field potential possessing only a mass term 
\cite{Kaup:1968zz,Friedberg:1986tq,Jetzer:1991jr,Liddle:1993ha}. Due to bounds on the maximal mass of this configurations 
they are usually referred to as {\it minimal boson stars}. In \cite{Colpi:1986ye} it was found that boson stars 
with masses of the order of astrophysical objects can be obtained including quartic self-interactions without collapsing into black holes. 
These studies were extended to include a 6th order term in the scalar field self-interaction potential \cite{kk1,kk2} as well
as exponential self-interaction potentials \cite{hartmann_riedel}.
Recently a very important application of this kind of configuration has been developed.
In \cite{Herdeiro:2014goa} the authors were able to construct Kerr black holes with scalar hair through a very particular 
interaction of rotating minimal boson stars and Kerr black holes. They have shown that once both configurations co-rotate 
synchronously it is possible to add hair onto the Kerr black holes. The mass of this new family of black holes is bounded by the 
mass of minimal boson stars. In a subsequently work \cite{Herdeiro:2015tia} the authors constructed and analyzed mass bounds and 
the size of the horizon of this kind of configurations when quartic self-interactions are included, in analogy to what was done in \cite{Colpi:1986ye} for boson stars. 
Boson stars have been constructed in several models including rotation, a cosmological constant and modified gravity theories as well as higher dimensional scenarios \cite{Astefanesei:2003qy, Brihaye:2009yr, Hartmann:2010pm, Brihaye:2013zha, Brihaye:2014bqa}. 
\\
Charged versions in both cases have attracted considerable attention as well, \cite{Kusenko:1997ad, Jetzer:1989av, Schunck:2003kk}. 
Once electromagnetic charge is introduced, the Coulomb electrostatic repulsion destabilizes 
large charged $Q$-balls \cite{Jetzer:1991jr, Kusmartsev:1990cr, Schunck:1999pm}. 
It was demonstrated for polynomial potentials \cite{Lee:1988ag} that there exists a maximum charge and size for these configurations (see
also \cite{Brihaye:2014gua} for a recent study). 
Attempts to construct large charged $Q$-balls including a V-shaped potential \cite{Arodz:2008nm,Kleihaus:2009kr} 
or even adding fermions \cite{Anagnostopoulos:2001dh} were given, but in all theses cases 
limitations on the maximum charge and size were found. Some advance in this direction was done in \cite{Tamaki:2014nua}
motivated by potentials belonging to the so--called Affleck-Dine mechanism. The formation of $Q$-balls in this context was also investigated \cite{Kasuya:1999wu}. 
\\
The self-interaction potential normally used to construct $Q$-balls and boson star is a 6th order polynomial potential
in the modulus of the complex scalar field.
Up to a rescaling, there is a unique choice of such a self-interaction potential which possesses -- in addition -- degenerate vacua. In this paper
we  construct $Q$-balls and boson stars in the context of such a potential with degenerate vacua.
We will report on the qualitative and quantitative properties of the solutions and compare our results with those 
given in \cite{Pugliese:2013gsa} for charged $Q$-balls
and boson stars with a scalar mass term only. 
\\
This paper is organized as follows: in Section 2 we give the model and discuss the possible soliton solutions 
(topological and non-topological, respectively).  In section 3, we discuss the case of one spatial dimension
and review what is known about solutions there. In Section 4, we will discuss solutions in three spatial dimensions. 
In this latter case, the non-topological solitons have to be constructed numerically both in flat as well as in curved space-time. In Section 5 we conclude and summarize our results.

\section{The model}
In the following we will discuss a scalar field model with a 6th order self-interaction potential whose action reads (in units such that
$\hbar=c\equiv 1$)
\begin{equation}
\label{action}
 S=\int \sqrt{-g} {\rm d}^{D+1} x \ {\cal L} = \int \sqrt{-g} {\rm d}^{D+1} x \ \left( -\partial_{\mu} \Phi \partial^{\mu} \Phi^* -  U(\vert\Phi\vert)\right)
\end{equation}
with scalar field potential
\begin{equation}
\label{potential}
U(\vert\Phi\vert) =  |\Phi|^2 \left(1- \vert\Phi\vert^2\right)^2  \ .
 \end{equation}
 $\mu=0,1,...,D$ and $g$ denotes the determinant of the metric
 tensor. In the following the signature of the metric is chosen to be $(-,+,...,+)$.  Note that there are already implicit rescalings here.
In principle, the value one in (\ref{potential}) should be replaced by an energy scale $\eta$ (which would correspond to the vacuum expectation value of the scalar field), but we can rescale $\Phi$ such that it is measured in units of this energy scale. Moreover, we have set the overall factor
in front of the potential to unity. This is nothing else but letting the mass of the scalar field $m^2$ be equal to unity. 
The potential (\ref{potential}) has three degenerate minima at $\vert\Phi\vert=\{-1,0,1\}$. This leads to interesting phenomena as far as 
soliton solutions are concerned. Let us mention that potentials with degenerate vacua also play an important r\^ole in the context
of an effective description of the confinement/deconfinement phase transition in Quantum Chromodynamics (see e.g. \cite{Brihaye:2012uw}
for the construction of soliton solutions in such an effective model).

Parametrizing the complex scalar field as
\begin{equation}
 \Phi=\vert\Phi\vert \exp(i\Gamma)   \ ,
\end{equation}
where, in general, both $\vert\Phi\vert\equiv \phi$ and $\Gamma$ can be functions of the $(D+1)$ coordinates, we are interested in the following two cases

\begin{enumerate}

\item $\Gamma=\omega t$: in this case the model possesses a conserved global $U(1)$ symmetry and with it a locally conserved Noether
current $j^{\mu}$, $\mu=0,1,...,D$
\begin{equation}
j^{\mu}
 = -i \left(\Phi^* \partial^{\mu} \Phi - \Phi (\partial^{\mu} \Phi)^*\right) \  \ {\rm with} \ \ \
\partial_{\mu} j^{\mu}=0  \ .
\end{equation}
The globally conserved Noether charge $Q$ of the system then reads
\begin{equation}
\label{charge}
 Q= -\int \sqrt{-g} \ j^0 \ d^{D}x  \  ,
\end{equation}
which can be interpreted as particle number. 
The choice of potential (\ref{potential}) allows for the existence of {\it non-topological solitons} called $Q$-balls \cite{Lee:1991ax} with
scalar field $\phi$ tending to zero at spatial infinity. These solutions exist for all $D \ge 1$. 
 
 \item $\Gamma\equiv 0$: in this case the scalar field is real and there exists no
 longer a $U(1)$ symmetry. The interesting solutions are quite different from the $Q$-balls: 
 instead of looking for configurations that tend to zero at infinity we focus on those that link the different vacua of 
 the potential $\phi(x=\pm \infty) = \{-1,0,1\}$. These solutions are \emph{topological solitons}. 
 Here the situation changes dramatically because the existence of such configurations is highly constrained by the number of spatial dimensions. 
 For the model given in \eqref{action} we only have topological solitons for $D=1$ and they are characterised by their topological charge:
$$
N = \frac{1}{2}\int_{-\infty}^{+\infty}\sqrt{-g}dx\;\frac{d\phi}{dx} = \frac{1}{2}\left[\phi(+\infty) - \phi(-\infty)\right] \ .
$$
The notion of soliton solutions in curved space-time is far from being easily generalised from the one in $(1+1)$ dimensions. The topological charge, however, can be directly extended (as given above) once the topological current in flat space
$$
j_{\textrm{top}}^\mu=\frac{1}{2}\epsilon^{\mu\nu}\partial_\nu \phi, \qquad \partial_\mu j^\mu_{\textrm{top}}=0
$$ 
is also covariantly zero: 
$$
\nabla_\mu j^{\mu}_{\textrm{top}} =\frac{1}{2}\epsilon^{\mu\nu}\nabla_\mu \varphi_\nu = 0 \Rightarrow \frac{1}{\sqrt{-g}}\partial_\mu\left(\sqrt{-g}j^\mu_{\textrm{top}}\right)=0.
$$
In relativistic theories in flat $(1+1)$--dimensional space-time the soliton energy $E$ (and with that, of course, the mass $M=E$) resembles the 
mass of a particle \cite{rajaraman}, specially due to its behaviour under transformations between inertial frames which move with speed $v$ relative
to each other: $E \rightarrow \frac{E}{\sqrt{1-v^2}}$. When gravity is also present, 
the intrinsic difficulty in General Relativity concerning the definition of a locally conserved energy makes 
this ``soliton-mass'' not well defined, and this is a serious issue in defining topological solitons in curved space-time.

\end{enumerate}

\section{Solitons in $(1+1)$ dimensions}
Scalar field models in $(1+1)$ dimensions have been studied extensively due to a variety of soliton solutions that exist in these
models. Though the choice of one spatial dimension might not be interesting from a physical application point of view,
these models possess rich mathematical structures.

\subsection{Topological solitons}

The model described by the action \eqref{action} in flat space-time with $D=1$ has interesting topological soliton solutions, 
namely kinks and anti-kinks, interpolating between the vacua of the potential \cite{Gani:2014gxa}. The static solutions read
\footnote{Note that there is a factor of $\sqrt{\frac{1}{2}}$ between the coordinates used in \cite{Gani:2014gxa} and those used in our paper.}
\begin{equation}
\label{phi6kink}
\phi(x)=\pm \sqrt{\frac{1}{2}\left[1\pm\tanh\left(\frac{x}{\sqrt{2}}\right)\right]} \ .
\end{equation}
The kinks reside either in the topological sector $N=1$ with energy $E=\pm \frac{1}{4}$ or in the topological sector
$N=2$ with energy $E=0$. Similarly, the anti-kinks reside in the topological sector $N=-1$ with energy $E=\pm \frac{1}{4}$ and
in the topological sector $N=2$ with energy $E=0$, respectively. For each topological sector these solutions are the minimum energy solutions \cite{Gani:2014gxa}.
\begin{figure}[h]
\begin{center}
{\includegraphics[width=10cm]{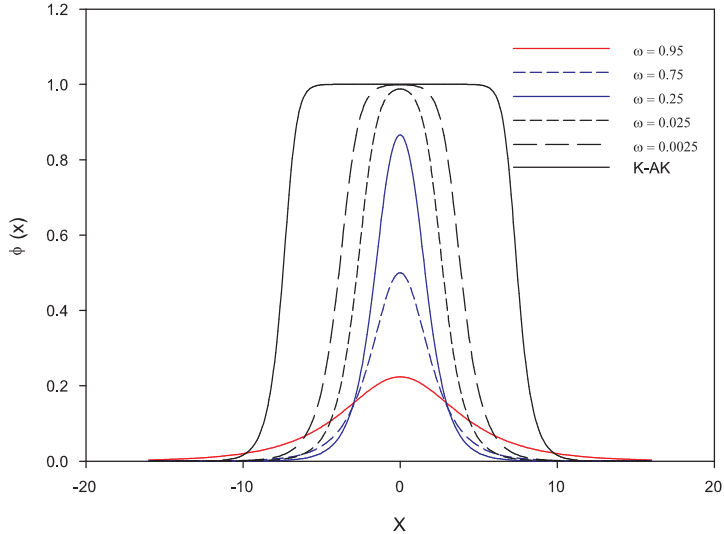}}
\end{center}
\caption{The profile of (\ref{approx}) describing approximately a kink-antikink solution in the $\phi^6$-potential (\ref{potential}) for
$a_2=-a_1=10$  (black solid). We also show the corresponding $Q$-ball solution (\ref{qball11}) for different values of $\omega$.
\label{kink_antikink} }
\end{figure}
The solitons can be translated and Lorentz-boosted to give the most general solution of the model
\begin{equation}
\label{phi6kink}
\phi(x,t)=\pm \sqrt{\frac{1}{2}\left[1\pm\tanh{\left(\sqrt{\frac{1}{2}}\frac{(x-a-vt)}{\sqrt{1-v^2}}\right)}\right]} \ ,
\end{equation}
where $v$ is a parameter which gives its speed of translation, 
and $a$ is another parameter, which can be interpreted as the position of the soliton.
Let us remark that these solutions are not strictly solitons because of their behaviour when they undergo collision processes 
(see the discussion in \cite{rajaraman} for the $\phi^4$--model). This model does not allow explicit 
multi-soliton solutions, i.e., solutions involving two kinks, or a kink and an anti-kink, etc. like, for instance, 
in the sine-Gordon model \footnote{This holds true for several \emph{integrable field theories}. 
The sine-Gordon model, in particular, also has kink-like solutions and for that reason we mention it here.}. In this latter model, the integrability, 
i.e. the existence of a Lax-Zakharov-Shabbat formulation, enables the construction of explicit 
$N$-soliton solutions \cite{laf_wojtek}. In fact, the construction of such more complex configurations 
in the $\phi^6$--model can only be done numerically, considering several approximations (see \cite{Gani:2014gxa} for details).
\\
For instance, when a kink and an anti-kink are sufficiently far apart so that they do not interact substantially, 
one can approximate the static solution by a product of a kink and an anti--kink solution as follows (see e.g. \cite{vachaspati}):
\\
\begin{equation}
\label{approx}
 \phi(x)=\left(\sqrt{\frac{1}{2}\left(1+\tanh\left(\frac{x}{\sqrt{2}}-a_1\right)\right)}\right)\cdot\left(\sqrt{\frac{1}{2}\left(1 - 
 \tanh\left(\frac{x}{\sqrt{2}}-a_2\right)\right)}\right) \ .
\end{equation}
\\
This solution is shown for $a_2=-a_1=10$ in Fig.\ref{kink_antikink}. 
Note that this configuration is unstable to decay to the solution $\phi(x)\equiv 0$ as it resides in the topologically trivial sector with topological
charge $N=0$.

\subsection{Non-topological solitons}
The introduction of a non-trivial phase allows for another kind of solution, as mentioned
above. For $D=1$ it is possible to write this non-topological soliton solution explicitly \cite{Lee:1991ax}:
\begin{equation}
\label{qball11}
\phi(x) =\frac{\sqrt{1-\omega^2}}{\sqrt{1+ \omega \cosh\left(2\sqrt{1-\omega^2}x\right)}} \ .
\end{equation}
This solution has a conserved Noether charge associated to the $U(1)$ symmetry. The
explicit expression for this charge is
\begin{equation}
Q=2\omega \int\limits_{-\infty}^{+\infty} \phi^2 \ {\rm d} x = 2\omega \sqrt{1-\omega^2} \ , 
\end{equation}
while the energy is $E\sim\omega^2\sqrt{1-\omega^2}$, i.e. $E\sim \omega Q$. 
For $\omega=0$ this solution is $\phi(x)\equiv 0$ and has vanishing Noether charge and energy.
The other extreme value is $\omega=1$. In this case, the Noether charge and energy is also vanishing. 
This solution is shown for several values of $\omega\in [0:1]$ in Fig.\ref{kink_antikink}. Increasing $\omega$ from zero the soliton
becomes broader in extend and at the same time its maximal value decreases. 

\section{Solitons in $(3+1)$ dimensions}
We would now like to study topological and non-topological solitons in $(3+1)$ dimensions. In this case, we can extend the model
to make it more general and couple the scalar field minimally to gravity and a $U(1)$ gauge field. 
The action $S$ of the field theoretical model that we are interested in the following reads
\begin{equation}
 S=\int \sqrt{-g} {\rm d}^4 x \left( \frac{R}{16\pi G} -\frac{1}{4}F_{\mu\nu}F^{\mu\nu}-D_{\mu} \Phi D^{\mu} \Phi^*
 - U(\vert\Phi\vert) \right)
\end{equation}
where $R$ is the Ricci scalar, $G$ denotes Newton's constant, $F_{\mu\nu}=\partial_{\mu} A_{\nu} - \partial_{\nu} A_{\mu}$ 
is the field strength tensor
of a $U(1)$ gauge field $A_{\mu}$ and $D_{\mu} \Phi = (\partial_{\mu} - i e A_{\mu})\Phi$ denotes the covariant
derivative with gauge coupling $e$.

The coupled field equations consist of the Einstein equations
\begin{equation}
\label{einstein}
 G_{\mu\nu}=8\pi G T_{\mu\nu}
\end{equation}
with $T_{\mu\nu}$ given by
\begin{eqnarray}
T_{\mu\nu} &=& g_{\mu\nu}{L}_M
-2 \frac{\partial {L}_M}{\partial g^{\mu\nu}}
=
    ( F_{\mu\alpha} F_{\nu\beta} g^{\alpha\beta}
   -\frac{1}{4} g_{\mu\nu} F_{\alpha\beta} F^{\alpha\beta})
\nonumber\\
&-& 
   \frac{1}{2} g_{\mu\nu} \left(     (D_\alpha \Phi)^* (D_\beta \Phi)
  + (D_\beta \Phi)^* (D_\alpha \Phi)    \right) g^{\alpha\beta}
  + (D_\mu \Phi)^* (D_\nu \Phi) + (D_\nu \Phi)^* (D_\mu \Phi) - 
    g_{\mu\nu}U(|\Phi|)   
  \label{tmunu}
\end{eqnarray}
and of the matter field equations
\begin{eqnarray}
\partial_\mu \left ( \sqrt{-g} F^{\mu\nu} \right) =
   \sqrt{-g} (- i e) \left(\Phi^* D^{\mu} \Phi - \Phi (D^{\mu} \Phi)^*\right)
     \ \ \ , \ \ \ D_\mu D^\mu \Phi - \frac{\partial U}{\partial \vert\Phi\vert^2}\Phi = 0
 . \label{feq_matter} \end{eqnarray}

\subsection{Topological solitons}
Static, finite energy solutions of pure scalar field models do not exist in dimensions $D\ge 2$. This
is Derrick's theorem (see e.g. \cite{Manton:2004tk}). Lifting the requirement of finite energy, a number of
topological solitons in pure scalar field models in $(3+1)$ dimensions exist. As an example let us mention a {\it domain wall}. This is a simple
extension of the static version of the solution given in (\ref{phi6kink}) to an additional two dimensions and reads
\begin{equation}
 \phi(x,y,z)=\pm \sqrt{\frac{1}{2}\left[1\pm\tanh\left(\frac{x}{\sqrt{2}}\right)\right]} \ .
\end{equation}
This describes a wall in the $y$-$z$-plane with finite energy per unit area, but, of course, infinite energy. While the energy
is infinite, these solutions play an important r\^ole in physical applications such as cosmology. 

On the other hand, finite energy solutions are possible
if one lifts the requirement of staticity. An example of this type of solution are $Q$-balls, which we will discuss
in the remainder of this paper. Note, however, that these solutions
are non-topological.

\subsection{Non-topological solitons}
In more than one spatial dimension these solutions have to be constructed numerically. In the following, we will construct these
solutions in three spatial dimensions assuming spherical symmetry. We also study the minimal coupling to gravity, in which case
the $Q$-balls are called {\it boson stars} in the literature, as well as to a $U(1)$ gauge field, i.e. gauging the $U(1)$ symmetry of the model.
For the numerical construction we adapt a collocation method with adaptive grid scheme \cite{colsys}.

\subsubsection{Ansatz, Equations of motion and boundary conditions}

In order to construct spherically symmetric, non-topological solitons, we use the following ansatz 
\begin{equation}
\label{metric}
 ds^2=-f(r) dt^2 + \frac{l(r)}{f(r)}\left[dr^2 + r^2 d\theta^2 + r^2 \sin^2\theta d\varphi^2\right]  \ 
\end{equation}
for the metric in isotropic coordinates and
\begin{equation}
\label{ansatz1}
\Phi(t,r)=\phi(r) e^{i\omega t}  \ , \ A_\mu d x^\mu = A(r) dt \ ,
\end{equation}
for the matter fields. Since the $U(1)$ symmetry is gauged in this case, the conserved Noether charge $Q$ can be interpreted as
the electric charge of the solution with $Q_e=e Q$. With our Ansatz, the explicit expression for the Noether charge reads
\begin{equation}
 Q= 8\pi \int\limits_{0}^{\infty} \frac{\sqrt{l^3}}{f^2} r^2 \left(\omega - e A\right)\phi^2 dr \ .
\end{equation}
Note that we still have the freedom of applying the following gauge transformation (with $\chi$ a constant)
\begin{equation}
 \Phi \rightarrow \Phi e^{-i\chi t} \ \ , \ \   A\rightarrow A + \frac{\chi}{e}  \ ,
\end{equation}
which can -- in principle -- be used to choose the scalar field real. 
However, we will not employ this here, but will use the gauge freedom to apply a constant shift to $A(r)$ such that
$A(0)=0$. 
The clear advantage is that we recover the limit $e\rightarrow 0$ naturally, because we do not ``gauge away'' the phase, which
has to be present in the limit of a global $U(1)$ symmetry. 

With our choice of potential (\ref{potential}) we have already introduced a rescaling into the system
by setting the mass of the scalar field equal to unity. We will use the abbreviation $\kappa=8\pi G$ in the
following. 
The explicit form of the coupled system of non-linear ordinary differential equations then reads
\begin{equation}
\label{phi_eq}
  \phi''=- \frac{r l' + 4l}{2rl} \phi'  -(\omega-e A)^2 \frac{l}{f^2} \phi 
  + \frac{l}{f} \phi\left(1-4\phi^2+ 3\phi^4\right)   \ ,
\end{equation}
\begin{equation}
\label{a_eq}
 A''=-\left(\frac{f'}{f} + \frac{l'}{2l} + \frac{2}{r}\right)A' + \frac{2l}{f} e(eA-\omega) \phi^2
\end{equation}
for the matter field functions and
\begin{equation}
 f'' = f'\left(-\frac{l'}{2l} + \frac{f'}{f} - \frac{2}{r}\right) + 2\kappa l 
\left(\frac{2(\omega-eA)^2 \phi^2}{f} - \phi^2(1-\phi^2)^2 \right)   \ ,
\end{equation}
\begin{equation}
\label{l_eq}
 l''=l'\left(\frac{l'}{2l} - \frac{3}{r}\right) + 4\kappa l^2 \left(\frac{(\omega-eA)^2 \phi^2}{f^2} 
 - \frac{\phi^2(1-\phi^2)^2}{f}\right)
\end{equation}
for the metric functions. The prime now and in the following denotes the derivative with respect to $r$. 
In the following we will study only $e > 0$ without loss of generality since the equations are invariant under $(e,A)\rightarrow (-e,-A)$.

The coupled set of non-linear ordinary differential equations (\ref{phi_eq}) - (\ref{l_eq}) has to be solved subject 
to appropriate boundary conditions. 
The regularity of the equations at the origin requires
\begin{equation}
\label{bc0}
 f'\vert_{r=0} = 0 \ \ , \ \ l'\vert_{r=0} = 0 \ \ , \ \ \phi'\vert_{r=0} = 0 \ \ , \ \ A'(0)=0 \ \ , \ \ A(0) = 0 \ ,
\end{equation}
where the last condition results from  the fixing of the residual gauge symmetry. 
The requirement of localized, finite energy, asymptotically flat solutions leads to 
\begin{equation}
\label{bcinf}
 f(r=\infty)=1 \ \ , \ \ l(r=\infty)=1 
\end{equation}
for the metric functions, while the matter functions have the following fall-off at infinity
\begin{equation}
 A(r \rightarrow \infty) \sim \mu + \frac{eQ}{r} \ \ , \ \  \phi(r\rightarrow\infty) \sim 
\frac{1}{r}\exp\left(- \Omega r\right) \ , \   \Omega=\sqrt{1-(\omega-e\mu)^2}
\end{equation}
where $\mu$ is a constant.

\subsubsection{Non-topological solitons in flat space-time: $Q$-balls}
Here we will discuss the properties of non-topological solitons in flat space-time, i.e. choosing $G=0$. The equations
then lead to $f(r)=l(r)\equiv 1$. 
These solutions have been discussed extensively for other type of potentials and the main goal of the following
will be to point out the differences to the cases studied before. We will first discuss the case of uncharged solutions, $e=0$,
in which case the model contains a global $U(1)$ symmetry and $Q$ can be interpreted as the number of scalar particles
forming the $Q$-ball. For $e > 0$ the $U(1)$ symmetry is gauged and the $Q$-ball possess an electric charge which is proportional
to the Noether charge. 

\begin{figure}[h]
\begin{center}
{\includegraphics[width=10cm]{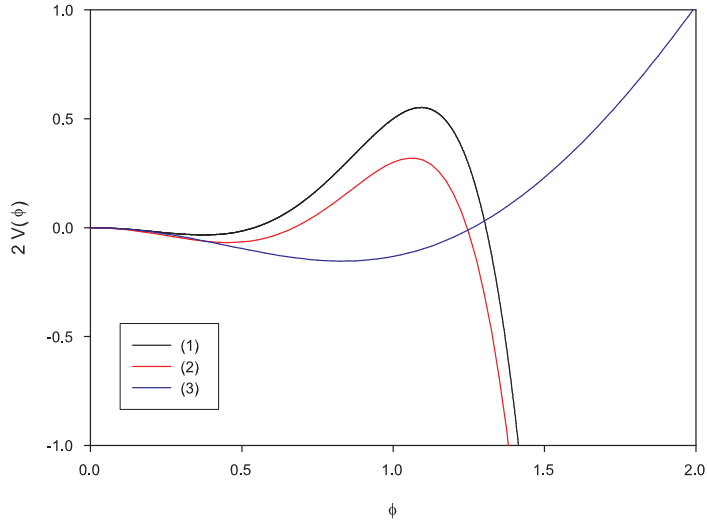}}
\end{center}
\caption{The potential $V(\phi)$ used in this paper for $\omega^2=0.5$ and $e\equiv 0$ (1). For comparison we also plot the corresponding
$V(\phi)$ for the potential used in \cite{Kleihaus:2005me} (2) as well as in \cite{Hartmann:2012gw} (3) for the same value of
$\omega^2$ and $e$. 
\label{potentials} }
\end{figure}

Let us remind the reader of the reasoning done in \cite{Volkov:2002aj} in order to understand the parameter restrictions.
The equation of motion for the scalar field can be rewritten as follows
\begin{equation}
 \frac{1}{2} \phi'^2 + \frac{1}{2}\left(\omega-eA\right)^2 \phi^2 -\frac{1}{2}U(\phi) = 
{\cal E} - 2\int\limits_0^r \frac{\phi'^2}{r} dr  \ ,
\end{equation}
where ${\cal E}$ is an integration constant. 
This describes the frictional motion of a particle with ``coordinate'' $\phi$ at ``time'' $r$ in an effective 
potential of the form
\begin{equation}
 V(\phi)= \frac{1}{2}\left(\omega-eA\right)^2 \phi^2 -\frac{1}{2}\left(\phi^2(1-\phi^2)^2\right) \ ,
\end{equation}
where we have inserted the explicit form of the potential (\ref{potential}). In Fig.\ref{potentials} we show this
potential $V(\phi)$ together with the corresponding potential for cases of other scalar field self-interaction potentials $U(\phi)$.
The potential (2) arises in the original discussion of $Q$-balls and boson stars \cite{Volkov:2002aj,Kleihaus:2005me}, where a 
potential was used that has has a single zero at $\phi\equiv 0$. The potential (3) is the exponential potential arising in supersymmetric
extensions of the Standard Model of Particle Physics. While the case (3) is quite different to the cases (1) and (2), the qualitative
features of (1) and (2) are similar. However, the main difference is that the potential (2) does not allow for
topological soliton solutions since the vacuum manifold is trivial in this case. 

The requirements for the existence of $Q$-balls are that $V''(\phi=0) < 0$ and $V(\phi) > 0$ for some $\phi\neq 0$ \cite{Coleman:1985ki}.
In our case this leads to the following conditions
\begin{equation}
(\omega-e\mu)^2 < 1 \ \ , \ \  \omega > 0   \ .
\end{equation}
Note that the latter condition appears only in the limit $e=0$. Furthermore, let us mention that the condition
$V''(0) < 0$ is nothing else but stating that the effective mass $m^2_{\rm eff}=m^2 - (\omega-e\mu)^2$ of the scalar particles forming the
$Q$-ball should be positive, where in our case $m^2\equiv 1$.
\\
\\

\paragraph{Uncharged $Q$-balls}
As stated above, uncharged $Q$-balls exist only in a limited range of the parameter $\omega \in [0,1]$.

\begin{figure}[h]
\begin{center}
{\label{v_6}\includegraphics[width=10cm]{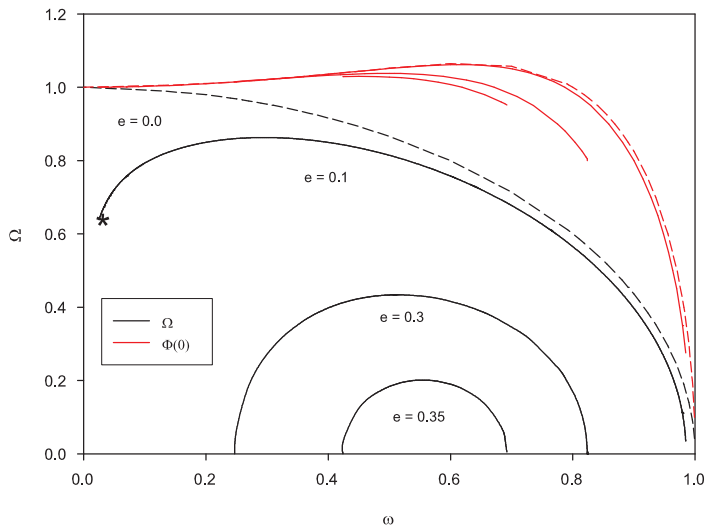}}
\end{center}
\caption{The values $\phi(0)$ (red) and $\Omega$ (black) as function of $\omega$ for uncharged (dashed) and charged (solid) $Q$-balls, respectively.
For $e=0.1$ we indicate the point at which our numerical analysis becomes unreliable by a star.
\label{omega_omega}
}
\end{figure}

\begin{figure}[h]
\begin{center}
{\label{v_6}\includegraphics[width=10cm]{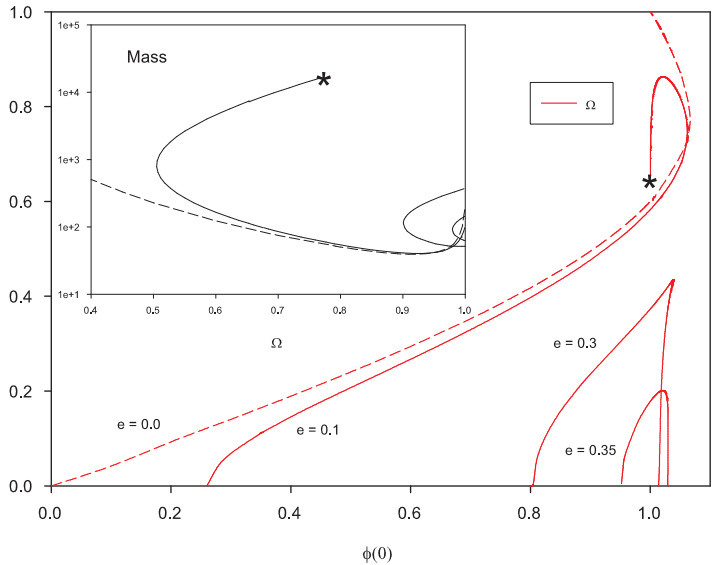}}
\end{center}
\caption{The value $\Omega=\sqrt{1-(\omega-e\mu)^2}$ as function of $\phi(0)$ for different values of $e$. For $e=0.1$ we indicate the point at which our numerical analysis becomes unreliable by a star.
\label{phi0_omega}
}
\end{figure}

In Fig. \ref{omega_omega} we show the value of $\phi(0)$ and $\Omega$ in dependence on
$\omega$. For $\omega\rightarrow 1$, the scalar field function tends to the zero function with $\phi(0)\rightarrow 0$. The reason
for this is that the effective mass $m_{\rm eff}\equiv \Omega=\sqrt{1 - \omega^2}$ of the scalar field becomes zero and hence a bound system of scalar particles 
is no longer possible. This is shown in Fig. \ref{phi0_omega}.  Decreasing $\omega$ from one we observe that $\phi(0)$ increases 
and has its maximal value $\phi(0)\approx 1.064$ 
at $\omega\approx 0.6$ and then decreases again to $\phi(0)=1$ at $\omega=0$, where $\Omega$ becomes equal to unity (see Fig. \ref{phi0_omega}).
Note that for $\omega=0$ the effective potential $V(\phi)$
becomes negative for any $\phi$ non-equal to one of the vacua of the potential $U(\phi)$. In this case, the scalar field
function behaves as $\phi(0)=1$ and $\phi(r > 0)=0$ corresponding to a step function. Hence, the kinetic energy of this solution diverges.
The regime close to this solution is sometimes called the {\it thin wall limit}. This 
is used referring to the 1-dimensional counterpart. Strictly speaking, in 3 spatial dimensions
the solutions are spherical balls the radius of which tends to zero in the
limit $\omega\rightarrow 0$. Hence, the model  in the limit $\omega\rightarrow 0$ allows for planar solutions with
infinite energy (the domain wall solutions mentioned above), but not for compact, spherically symmetric solutions. 

\begin{figure}[h]
\begin{center}
{\label{v_6}\includegraphics[width=10cm]{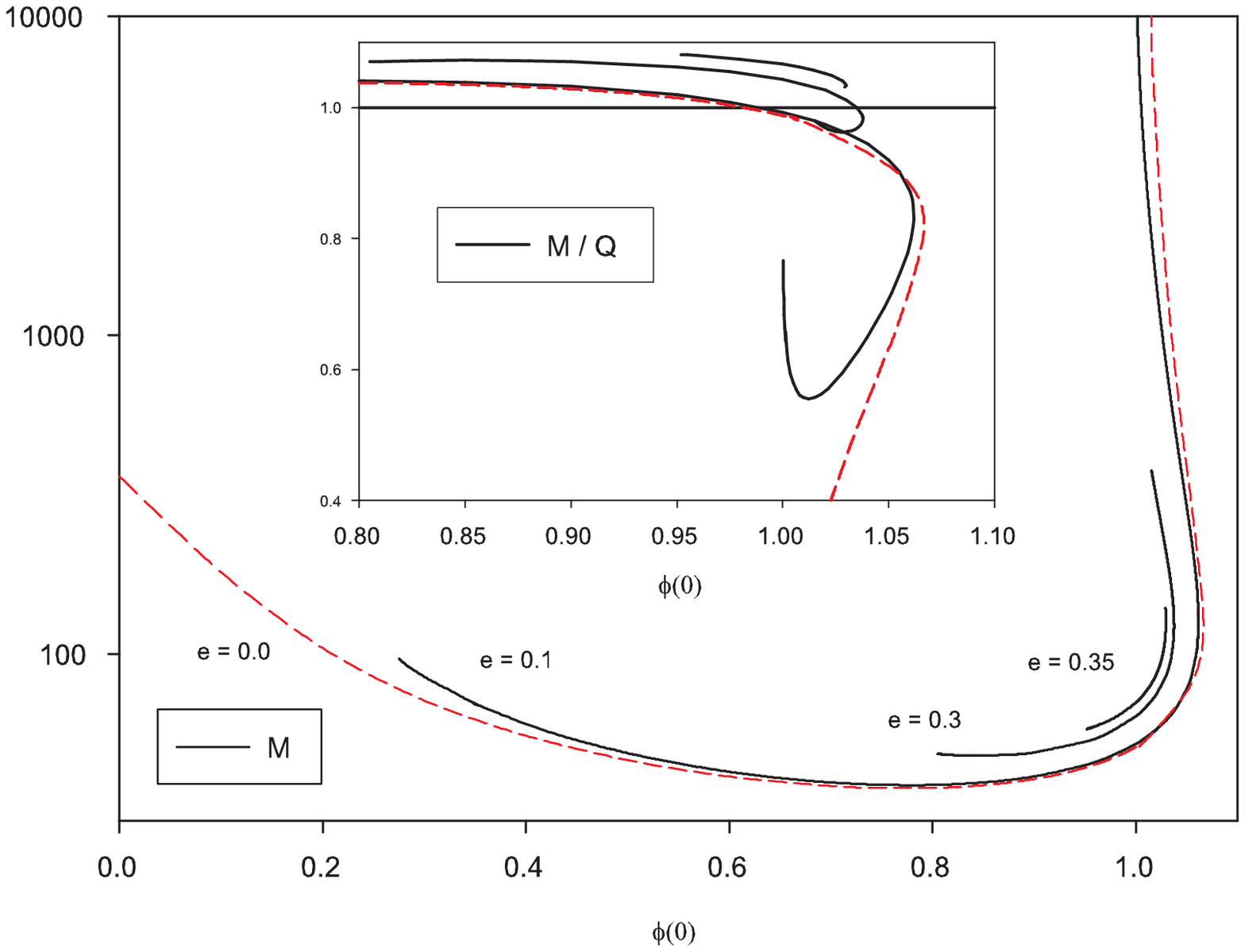}}
\end{center}
\caption{We show the mass of $Q$-balls  as function of $\phi(0)$ for different values of $e$.
The subfigure shows the ratio of mass over Noether charge $M/Q$ as function of $\phi(0)$ close to $\phi(0)=1$. 
For $M/Q < 1$ the $Q$-balls are classically stable to decay into $Q$ individual scalar bosons with mass $m^2\equiv 1$, otherwise
unstable. 
\label{phi0_mass}
}
\end{figure}

In both limits, i.e. $\omega\rightarrow 0$ and $\omega\rightarrow 1$ the mass of the $Q$-ball diverges. This is shown in Fig.\ref{phi0_mass}, where
we give the mass $M$ as function of $\phi(0)$. We observe that
the mass is minimal for $\phi(0) \approx 0.8$. 
In Fig. \ref{phi0_mass} we also show the mass $M$ and the ratio $M/Q$ as function of $\phi(0)$. This latter ratio is interesting
in order to compare the mass of the $Q$-ball $M$ with the mass of $Q$ individual scalar bosons of mass $m^2=1$. 

The quantity $M/Q-1$ is negative only for the solutions corresponding
to $\omega < 0.81$ (i.e. $\phi(0) > 0.98$). Hence, we would only expect the solutions in the thin wall limit
to be classically stable against decay into $Q$ individual bosons with mass $m^2\equiv 1$. 

\paragraph{Charged $Q$-balls}

Charged $Q$-balls with non-polynomial potentials have been studied previously:
in \cite{Arodz:2008nm} for a non-differentiable sign-Gordon potential 
and in \cite{Brihaye:2014gua} for a exponential potential arising from gauge-mediated supersymmetry breaking
of the minimal supersymmetric extension of the standard model of particle physics. 

\begin{table}[h]
\centering
\begin{tabular}{|c|c|c|}
\hline
$e$ & $\phi(0)_{\rm min}$ & $\phi(0)_{\rm max}$\\
\hline
0.00 & 0.00 & 1.000\\
0.10 & 0.26 & 1.000\\
0.30 & 0.81 & 1.014\\
0.35 & 0.95 & 1.029 \\
\hline
\end{tabular}
\caption{The minimal value $\phi(0)_{\rm min}$ as well as the maximal value $\phi(0)_{\rm max}$ of the central value of $\phi(0)$ for a given value of $e$ for charged $Q$-balls, see also Fig.\ref{phi0_omega}.}
\label{table1}
\end{table}

As expected from the reasoning above, the pattern of solutions is quite different for $e > 0$.
In particular, charged $Q$-balls do not exist on
the full interval $\omega \in [0,1]$, but only on a smaller interval
$\omega \in [\omega_m, \omega_M]$ whose bounds depend  on the value of $e$. This is shown
in  Fig. \ref{omega_omega}, where we give the value of $\Omega$ as function of $\omega$, as well as in 
Fig. \ref{phi0_omega}, where we give the value of $\Omega$ as function of $\phi(0)$. 
With increasing $e$ the interval for which charged $Q$-balls exist decreases both in $\omega$ as well
as in $\phi(0)$. The reason is that the higher the value of $e$ the stronger is the electromagnetic repulsion
and hence charged $Q$-balls cannot have arbitrarily large central density. Or stated differently:
charged $Q$-balls exist only when the  scalar field is dense enough to 
compensate the electrostatic repulsion of its constituents. At the approach of the values $\omega_{m}$, $\omega_M$
the quantity $\Omega$ tends to zero. Accordingly, the solutions cease to be exponentially localized. In Table \ref{table1}, we give the 
minimal and maximal value of $\phi(0)$ for fixed values of $e$. The extrapolation of this data suggests that
for $e \gtrsim 0.38$, where $\phi_{\rm min}(0)=\phi_{\rm max}(0)$, charged $Q$-balls do no longer exist. 

We also observe that the stable branch of solutions connected to $\omega_M$ becomes smaller in extend. 
Hence, for $ 0.38  \gtrsim e > 0.35$ charged $Q$-balls exist, but they are no longer classically stable. This is seen in Fig.\ref{phi0_mass}.

The values of $e$ that we discussed above provide a good representation of the pattern of solutions.
We observe that all $Q$-balls have $\mu < 0$ for all solutions constructed that the chemical potential is always negative
and that the quantity $\omega - e \mu$ is positive definite.

Let us now compare this to the case of charged $Q$-balls in an exponential potential arising in the minimal extension of the
Standard Model, which has been discussed in \cite{Brihaye:2013zha}. The first thing to note is that already the uncharged case, i.e. the case $e\equiv 0$ 
is different. While uncharged $Q$-balls in an exponential potential can have arbitrarily large central value $\phi(0)$, $Q$-balls in the
potential studied in this paper exist only for $\phi(0)\leq 1.064$. Moreover, in the interval $\phi(0)\in [1:1.064]$ two solutions
with same central value $\phi(0)$ exist which, however, have different values of $\Omega$ and hence exponential fall-off of the scalar
field function. This is not possible in the case of an exponential potential.

\subsubsection{Non-topological solitons in curved space-time: boson stars}
In the following we will use the abbreviation $\kappa=8\pi G$ and restrict ourselves to the comparison between
the cases $\kappa=0$ and $\kappa=0.2$. 

\paragraph{Uncharged boson stars}
We will first discuss the case of uncharged boson stars and investigate the influence of gravity on the structure of solutions.
In Fig. \ref{null_charge_1} we show the frequency $\omega$ as function of $\phi(0)$ for $\kappa=0.2$ and -- for comparison -- also give the
corresponding curve for $\kappa=0$. As is apparent from this figure, the inclusion of gravity now allows
for arbitrarily large values of the central value $\phi(0)$. We only show the curve up to $\phi(0)=3$, but it continues to $\phi(0)\rightarrow \infty$.
In the limit of large $\phi(0)$ the value of the metric function $f(r)$ at the origin, $f(0)$, tends to zero making the solution
singular in this limit. The qualitative behaviour of the mass $M$ and the Noether charge $Q$ in dependence on either $\phi(0)$ (see Fig.\ref{null_charge_2}) 
or $\omega$ is very similar to the case with other potentials studied before:

\begin{figure}[h!]
\begin{center}
\mbox{
\subfigure[][]{
\includegraphics[height=.3\textheight, angle =0]{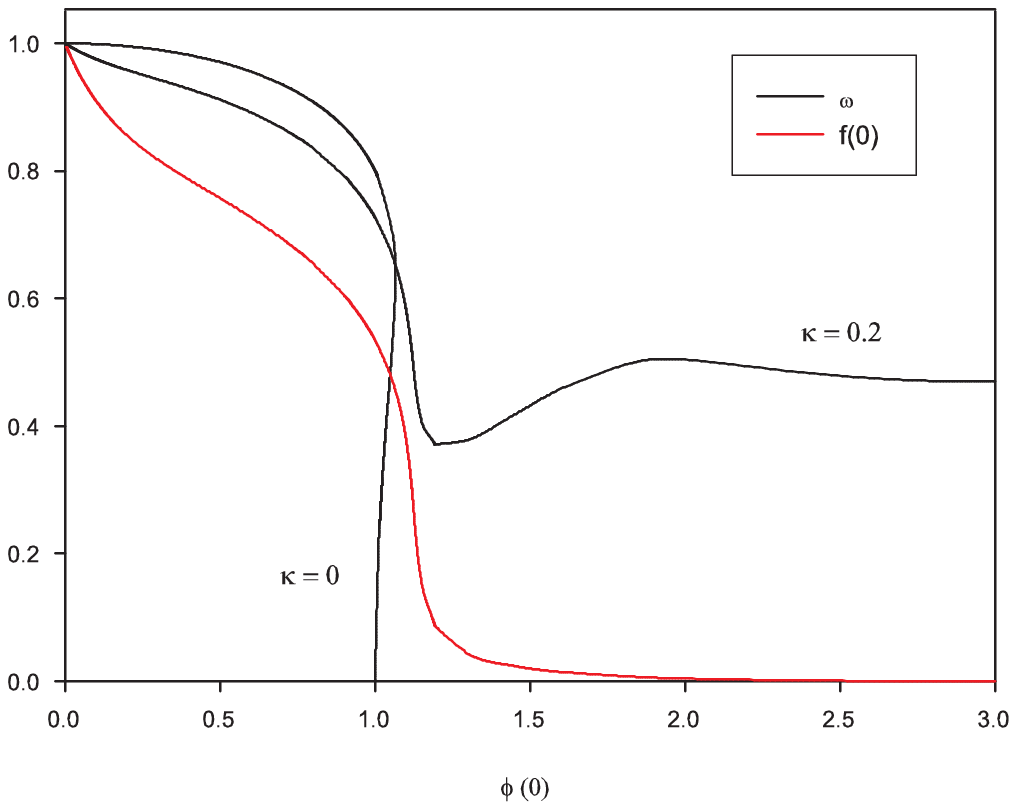}
\label{null_charge_1}
}
\subfigure[][]{
\includegraphics[height=.3\textheight, angle =0]{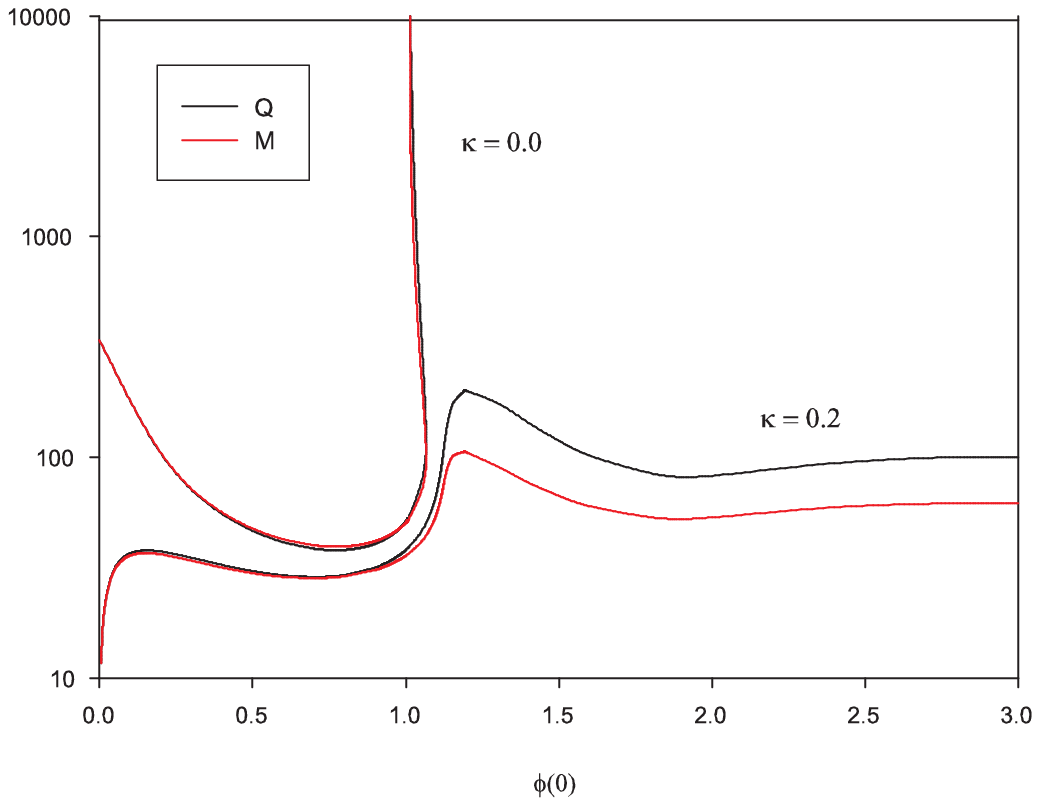}
\label{null_charge_2}
}
}
\end{center}
\vspace{-0.5cm}
\caption{We show the frequency $\omega$ and the value of the metric function $f(r)$ at the origin, $f(0)$ (a) as well as the charge $Q$ and mass $M$ (b) as function of $\phi(0)$ for uncharged boson stars
with $\kappa=0.2$. For comparison we give also the corresponding data for the $Q$-balls with $\kappa=0$. 
\label{null_charge}
}
\end{figure}

The gravitational interaction  
influences substantially the  two limits   $\phi(0) \ll 1$ and $\phi(0) \gg 1$.  
The limit $\phi(0) \to 0$ corresponds to $\omega \to 1$ (irrespectively of $\kappa$). In this limit -- and this is the main
difference to the $\kappa=0$ case -- both $M$ and $Q$ tend monotonically to zero.
Decreasing $\omega$ (corresponding to increasing $\phi(0)$) the solutions exist only down to a minimal and finite
value of $\omega$. Hence, the $\omega\rightarrow 0$ limit is not reachable in curved space-time. Moreover, several
branches of the solutions exist and form a spiral typical for boson star solutions. 

\paragraph{Charged boson stars}
Charged boson star solutions in a scalar field model without self-interaction have been studied in detail in \cite{Pugliese:2013gsa}, while
charged boson stars in a V-shaped potential and exponential scalar potential, respectively, have been studied in 
\cite{Kleihaus:2009kr} and \cite{Brihaye:2013zha}.

\begin{figure}[h]
\begin{center}
{\label{v_6}\includegraphics[width=10cm]{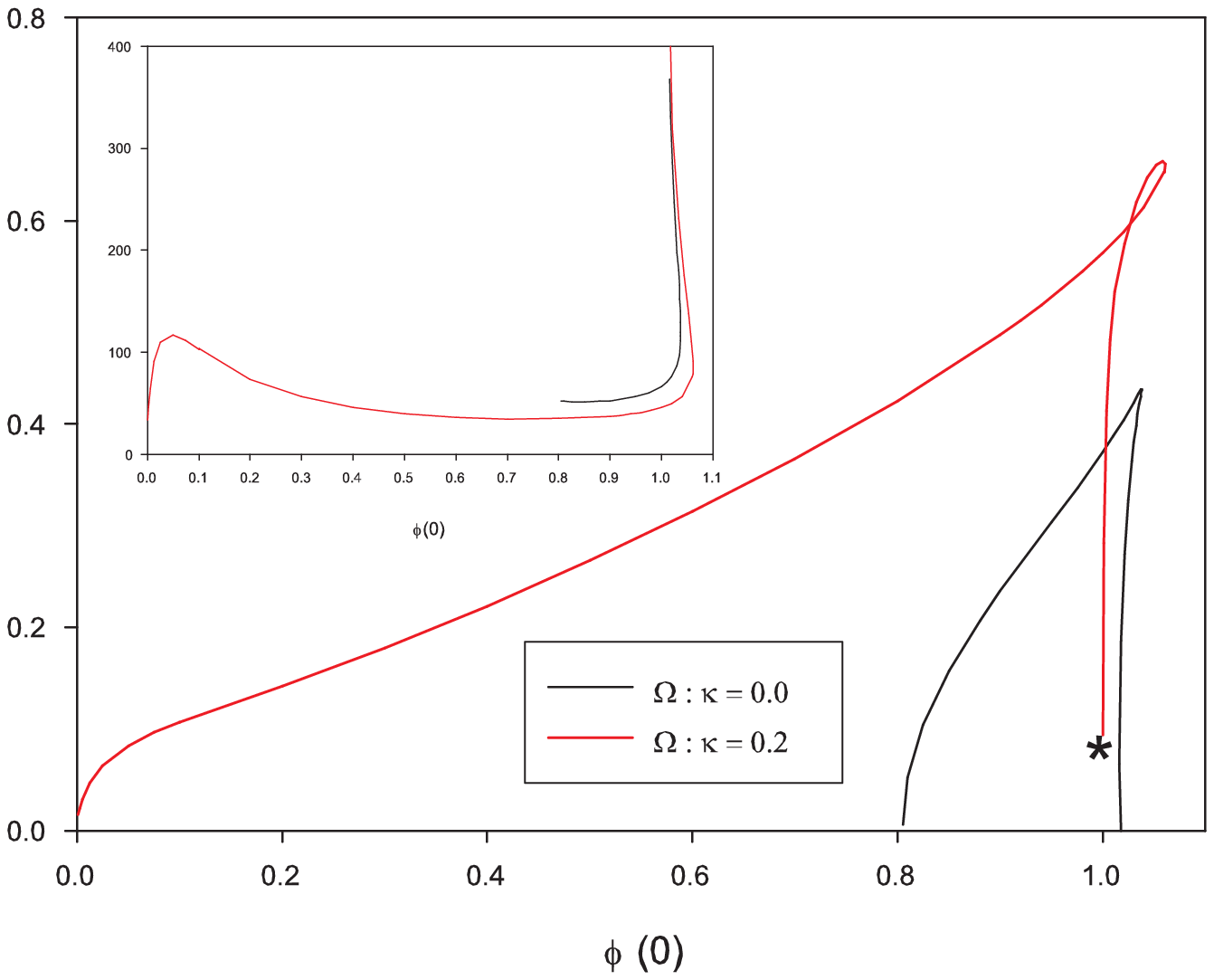}}
\end{center}
\caption{We show the parameter $\Omega$ as function of $\phi(0)$ for charged boson stars
with $\kappa=0.2$ (red) and charged $Q$-balls (black)
for $e=0.3$. The corresponding mass $M$ of the solutions is shown in the subfigure.
\label{gravity_nogravity_1}
}
\end{figure}

\begin{figure}[h]
\begin{center}
{\label{v_6}\includegraphics[width=10cm]{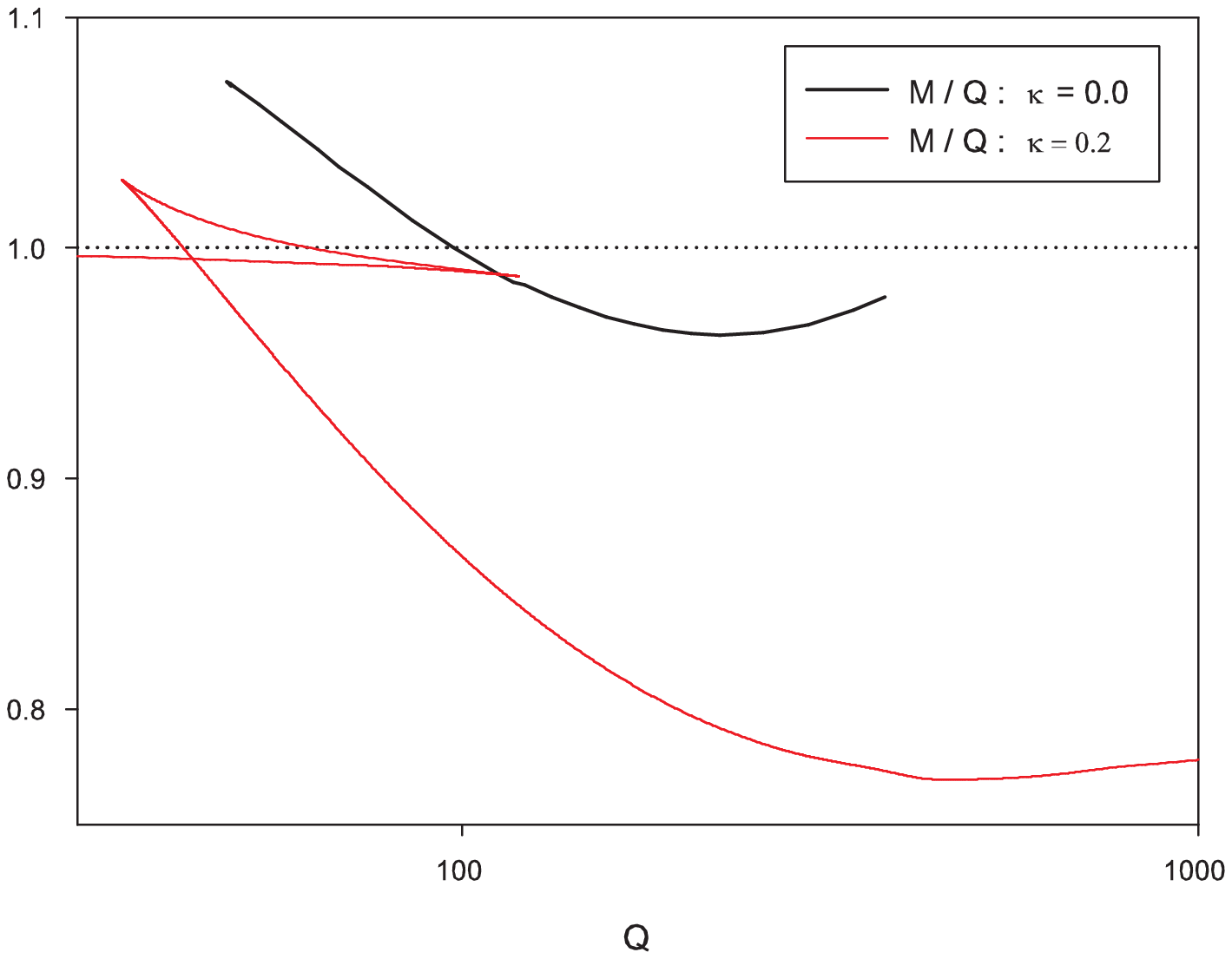}}
\end{center}
\caption{We show the ratio of the mass $M$ and the charge $Q$ as function of $Q$ for charged boson stars
with $\kappa=0.2$ (red) and $Q$-balls (black)
for $e=0.3$.
\label{gravity_nogravity_3}
}
\end{figure}

In the following, we will present our results for $e=0.3$, but we checked that the qualitative features are similar 
for other values of $e$. 
Our results are given in Fig.\ref{gravity_nogravity_1} and Fig.\ref{gravity_nogravity_3}, where we present the
data for gravitating and charged boson stars (red lines) and compare them to that of charged $Q$-balls (black lines).

We observe that the charged boson stars exist in the limit $\phi(0) \to 0$ and that the mass
$M$, the particle number $Q$ and the parameter $\Omega$ 
tend to zero in the limit $\phi(0) \to 0$.
In this limit, the gravitational attraction manages to compensate the electrostatic repulsion
even for boson stars that are composed of a small number of scalar quanta. 

Moreover, contrasting to the uncharged case, charged $Q$-balls and boson stars cannot be constructed 
for large values of the central value $\phi(0)$. When this 
parameter approaches a critical value close to unity,
a thin wall limit is approached and both the mass and the charge diverge. The values $\Omega$ and $f(0)$ both approach zero in this limit.

As expected, the parameter range in which boson stars are stable is larger than in the case  of $Q$-balls 
and the binding energy is stronger. This is seen in Fig. \ref{gravity_nogravity_3} where the ratio $M/Q$
is given as function of the particle number $Q$. For boson stars, this ratio tends to one in the limit $\phi(0) \to 0$.

\paragraph{Comparison with charged boson stars without self-interaction}

The construction of charged boson star in a scalar field model without self-interaction, i.e. only with a mass term for the scalar field
was elaborated in detail recently \cite{Pugliese:2013gsa}. 
Here we want to compare the properties of the charged boson stars in that model with those studied in this paper.
We will choose $\kappa= 0.2$ and $e=0.3$ in the following to point out the main differences.

\begin{figure}[h]
\begin{center}
{\label{v_6}\includegraphics[width=12cm]{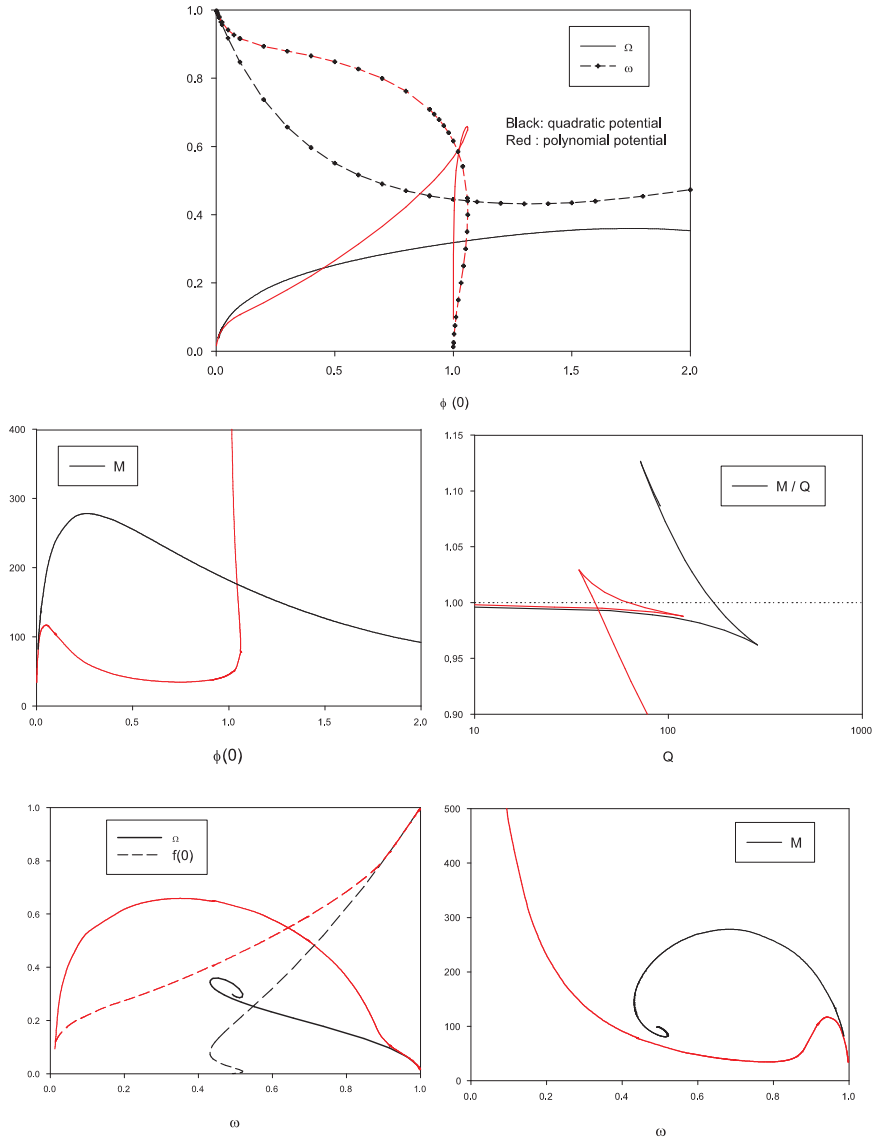}}
\end{center}
\caption{We compare several parameters characterizing the 
boson stars for $\kappa = 0.2$, $e = 0.3$ for a model with mass potential $U(\phi)=m^2\phi^2=\phi^2$ only
(black) and a model with the polynomial potential (\ref{potential}) (red).
\label{comparaison_potential}
}
\end{figure}

Several parameters characterizing the solutions are given in Fig. \ref{comparaison_potential}.
We observe that for $\phi(0) \to 0$ the features are qualitatively similar.
In particular, for $\phi(0) \geq 0$ the solutions smoothly approach flat space-time with $\omega \to 1$.
This result can be easily understood by noticing that, for such solutions, the scalar field 
remains confined around the local minimum occuring at $\phi=0$.

The properties of the solutions of the two models, however, differs significantly for large values of the central density $\phi(0)$.
In particular,  minimal boson stars are not limited by a thin wall limit  
when $\phi(0)$ becomes large. The mass and the charge remain finite in this limit while 
the metric  function $f(r)$ at the origin, $f(0)$, slowly approaches zero.
This results is related to an increase of the energy density of matter around the origin. 

The presence of the scalar self-interaction also influences the stability of the solutions.
With the set of parameters choosen, the boson star without self-interaction is stable for (essentially) all values
of the charge. By contrast the solutions composed of a self-interacting scalar field become unstable in the thin wall
limit, where mass and charge become large.

\clearpage
\section{Conclusions}
In this paper we have studied a scalar field model with a potential that contains three degenerate mimima.
In (1+1) dimensions this model allows for topological solitons and non-topological solitons when the scalar field is
chosen real and complex, respectively. The topological solitons are the analogues of (anti-)kink solutions and the analytic
solutions have been given in \cite{Gani:2014gxa}. Just like the $\phi^4$-kinks these
can be trivially extended to (3+1) dimensions, where they correspond to domain walls. The non-topological solutions
existing in this model are $Q$-balls with a globally conserved Noether charge. In (1+1) dimensions the
solutions can be given explicitly \cite{Lee:1991ax}, while in more than one spatial dimension they have to be constructed numerically.
One of the aims of this paper is to present our results on the construction of these $Q$-ball solutions in (3+1) dimensions. In contrast to 
the topological solitons, these non-topological solitons have finite energy. We have also extended our results to include 
a minimal coupling to gravity and electromagnetism, respectively. In the former case, the solutions are also often
called {\it boson stars} since they represent compact, gravitating objects that are made up of scalar bosons. 
In certain cases, our solutions possess a so-called {\it thin wall limit} which appears in the limit in which the
scalar field becomes real. 
\\
It is interesting to note that in (1+1) dimensions the approximate kink-antikink solution resembles the corresponding $Q$-ball solution.
Both reside in the topologically trivial sector. It is well known that $Q$-balls typically possess two branches of solutions when
plotting the mass $M$ over the Noether charge $Q$. The lower energy branch is connected to the Minkowski vacuum with $M=Q=0$ 
(and as such is linearly stable), 
while the second branch is the branch of unstable
$Q$-ball solutions. We conjecture that the kink-antikink solution is, in fact, the end point of the second
branch of unstable $Q$-ball solutions. Or to state it differently: the two branches of $Q$-ball solutions connect the linearly
stable Minkowski vacuum to the kink-antikink solutions. Due to catastrophe theory, which states that the number of instabilities changes
only at cusps where two branches of solutions meet, we can hence also conjecture -- not surprisingly -- 
that the kink-antikink solution is linearly unstable.
It is interesting to see whether these features persist for other type of potentials such as the $\phi^8$-potential in which
kink solutions have been constructed \cite{Gani:2015cda} as well as the exponential potential appearing in gauge-mediated supersymmetry breaking 
\cite{Copeland:2009as}. 

Finally, let us mention that the obtained results are interesting in order to understand better the dynamics of the system.
Kink-antikink collisions in integrable models such as the sine-Gordon model reveal the true {\it solitonic} nature of the latter
in the sense that no annihilation takes place in the collision and the solitons keep their initial velocities after the collision.
It was already observed some time ago that kink-antikink collisions in non-integrable models are in contrast inelastic. 
This has been studied thoroughly in the $\phi^4$-model \cite{Moshir:1981ja,Campbell:1983xu,Campbell:1986mg,Campbell:1986nu,Anninos:1991un}
and it was attributed to the fact that the $\phi^4$-model possesses internal modes that can be excited, which the sine-Gordon model does not possess. 
Kink-antikink collisions in the $\phi^6$ case have been studied recently with view to this
point \cite{Hoseinmardy:2010zz,Dorey:2011yw} and it was found that the $\phi^6$ model behaves in many aspects similar to the $\phi^4$-model.
We would like to emphasize, though, that the $\phi^6$-model is different from the $\phi^4$-model in the sense that only the former 
possesses $Q$-balls solutions. Hence, it is conceivable that in the collision of $\phi^6$ kinks and antikinks 
(which should both asymptote to zero) $Q$-balls get formed.
The fact that this is not seen in the papers mentioned above is likely due to the fact that only linear perturbations about the kink and anti-kink, respectively,
have been taken into account. We conjecture that when allowing for non-linear perturbations, the $Q$-balls will form. This is currently under investigation.

\section*{Acknowledgments}

A.C. work is supported by FONDECYT project N\textordmasculine3150157. G.L. acknowledges the support by the International Cooperation Program CAPES-ICRANet financed by CAPES - Brazilian Federal Agency for Support and Evaluation of Graduate Education within the Ministry of Education of Brazil.


\clearpage
\begin{thebibliography}{99}

\bibitem{Polchinski:1998rq} 
  J.~Polchinski, {\it String theory. Vol. 1: An introduction to the bosonic string}, (1998).

\bibitem{Tong:2009np} 
  D.~Tong, {\it String Theory},
  arXiv:0908.0333 [hep-th].
 
  

\bibitem{Fon9} M.J.~Duff, B.E.W.~Nilsson and C.N.~Pope, Phys.\ Rept \textbf{130}, 1-142 (1986).

\bibitem{Freedman:2012zz} 
  D.~Z.~Freedman and A.~Van Proeyen, {\it Supergravity}, (2012).
 

\bibitem{Joyce:2014kja} 
  A.~Joyce, B.~Jain, J.~Khoury and M.~Trodden,
  Phys.\ Rept.\  {\bf 568}, 1 (2015).
  
\bibitem{Starobinsky:1980te} 
  A.~A.~Starobinsky,
  Phys.\ Lett.\ B {\bf 91}, 99 (1980).
 
  
\bibitem{atlas} G. Aad et al. [ATLAS Collaboration], Phys. Lett. B \textbf{716}, 1 (2012); S. Chatrchyan et al. [CMS Collaboration], Phys. Lett. B \textbf{716}, 30 (2012).

\bibitem{Manton:2004tk} 
  N.~S.~Manton and P.~Sutcliffe, {\it Topological solitons}, Cambridge University Press (2004).
 
 
\bibitem{zabusky} 
N.~J~.Zabusky and M.~D.~Kruskal,
Phys. \ Rev. \ Lett. {\bf 15}, 240 (1965).


\bibitem{laf_wojtek} L.~A.~Ferreira and W.~J.~Zakrzewski,
  JHEP {\bf 0709} (2007) 015.


  
  
  
  
\bibitem{Coleman:1985ki} 
  S.~R.~Coleman,
  Nucl.\ Phys.\ B {\bf 262}, 263 (1985)
  [Nucl.\ Phys.\ B {\bf 269}, 744 (1986)].
  
  \bibitem{Lee:1991ax}
  T.~D.~Lee and Y.~Pang,
  Phys.\ Rept.\  {\bf 221} (1992) 251.
  
  
  
\bibitem{Volkov:2002aj} 
  M.~S.~Volkov and E.~W\"ohnert,
  Phys.\ Rev.\ D {\bf 66}, 085003 (2002).
 
  
\bibitem{Kleihaus:2005me} 
  B.~Kleihaus, J.~Kunz and M.~List,
  Phys.\ Rev.\ D {\bf 72}, 064002 (2005).
  
  
\bibitem{Kusenko:1997zq} 
  A.~Kusenko,
  Phys.\ Lett.\ B {\bf 405}, 108 (1997).
  
\bibitem{Kusenko:1997ad} 
  A.~Kusenko,
  Phys.\ Lett.\ B {\bf 404}, 285 (1997).
  
  
\bibitem{Hartmann:2012gw}
  B.~Hartmann and J.~Riedel,
  Phys.\ Rev.\ D {\bf 87} 4, 044003 (2013).
 

 
\bibitem{Kaup:1968zz} 
  D.~J.~Kaup,
  Phys.\ Rev.\  {\bf 172}, 1331 (1968).
  
\bibitem{Friedberg:1986tq} 
  R.~Friedberg, T.~D.~Lee and Y.~Pang,
  Phys.\ Rev.\ D {\bf 35}, 3658 (1987).
  
\bibitem{Jetzer:1991jr} 
  P.~Jetzer,
  Phys.\ Rept.\  {\bf 220}, 163 (1992).
  
  
\bibitem{Liddle:1993ha} 
  A.~R.~Liddle and M.~S.~Madsen,
  Int.\ J.\ Mod.\ Phys.\ D {\bf 1}, 101 (1992).
  
\bibitem{Colpi:1986ye} 
  M.~Colpi, S.~L.~Shapiro and I.~Wasserman,
  Phys.\ Rev.\ Lett.\  {\bf 57}, 2485 (1986).
  
 \bibitem{kk1} B. Kleihaus, J. Kunz and M. List, Phys. Rev. D {\bf 72} (2005), 064002.
\bibitem{kk2} B. Kleihaus, J. Kunz, M. List and I. Schaffer, Phys. Rev. D {\bf 77} (2008), 064025. 
  

 
\bibitem{hartmann_riedel} 
 B.~Hartmann and J.~Riedel,
  Phys.\ Rev.\ D {\bf 86} (2012) 104008.
  
\bibitem{Herdeiro:2014goa} 
  C.~A.~R.~Herdeiro and E.~Radu,
  Phys.\ Rev.\ Lett.\  {\bf 112}, 221101 (2014)
  
  
\bibitem{Herdeiro:2015tia} 
C.~A.~R.~Herdeiro, E.~Radu and H.~Rúnarsson,
  Phys.\ Rev.\ D {\bf 92} (2015) 8,  084059

 
\bibitem{Astefanesei:2003qy} 
  D.~Astefanesei and E.~Radu,
  Nucl.\ Phys.\ B {\bf 665}, 594 (2003)
 
  
\bibitem{Brihaye:2009yr} 
  Y.~Brihaye, T.~Caebergs, B.~Hartmann and M.~Minkov,
  Phys.\ Rev.\ D {\bf 80}, 064014 (2009).
  
  
\bibitem{Hartmann:2010pm} 
  B.~Hartmann, B.~Kleihaus, J.~Kunz and M.~List,
  Phys.\ Rev.\ D {\bf 82}, 084022 (2010).
  
  
\bibitem{Brihaye:2013zha} 
  Y.~Brihaye and J.~Riedel,
  Phys.\ Rev.\ D {\bf 89}, no. 10, 104060 (2014).
  
\bibitem{Brihaye:2014bqa} 
  Y.~Brihaye, B.~Hartmann and J.~Riedel,
  Phys.\ Rev.\ D {\bf 92}, no. 4, 044049 (2015).
  
  
  
 
  
  
  
\bibitem{Jetzer:1989av} 
  P.~Jetzer and J.~J.~van der Bij,
  Phys.\ Lett.\ B {\bf 227}, 341 (1989).
  
  
\bibitem{Schunck:2003kk} 
  F.~E.~Schunck and E.~W.~Mielke,
  Class.\ Quant.\ Grav.\  {\bf 20}, R301 (2003).
  
  
\bibitem{Kusmartsev:1990cr} 
  F.~V.~Kusmartsev, E.~W.~Mielke and F.~E.~Schunck,
  Phys.\ Rev.\ D {\bf 43}, 3895 (1991).
  
\bibitem{Schunck:1999pm} 
  F.~E.~Schunck and E.~W.~Mielke,
  Gen.\ Rel.\ Grav.\  {\bf 31}, 787 (1999).
  
\bibitem{Lee:1988ag} 
  K.~M.~Lee, J.~A.~Stein-Schabes, R.~Watkins and L.~M.~Widrow,
  Phys.\ Rev.\ D {\bf 39}, 1665 (1989).
 
 
  
\bibitem{Brihaye:2014gua}
  Y.~Brihaye, V.~Diemer and B.~Hartmann,
  Phys.\ Rev.\ D {\bf 89} (2014) 084048
  [arXiv:1402.1055 [gr-qc]].
 
 
\bibitem{Arodz:2008nm} 
  H.~Arodz and J.~Lis,
  Phys.\ Rev.\ D {\bf 79}, 045002 (2009)
  
  \bibitem{Kleihaus:2009kr}
  B.~Kleihaus, J.~Kunz, C.~L\"ammerzahl and M.~List,
  Phys.\ Lett.\ B {\bf 675} (2009) 102
 
\bibitem{Anagnostopoulos:2001dh} 
  K.~N.~Anagnostopoulos, M.~Axenides, E.~G.~Floratos and N.~Tetradis,
  Phys.\ Rev.\ D {\bf 64}, 125006 (2001).
 
  
  

  
  
\bibitem{Tamaki:2014nua} 
  T.~Tamaki and N.~Sakai, {\it Gauged Q-balls in the Affleck-Dine mechanism}, 
  arXiv:1401.0996 [hep-th].
  
\bibitem{Kasuya:1999wu} 
  S.~Kasuya and M.~Kawasaki,
  Phys.\ Rev.\ D {\bf 61}, 041301 (2000).
  
\bibitem{Pugliese:2013gsa} 
  D.~Pugliese, H.~Quevedo, J.~A.~Rueda H. and R.~Ruffini,
  Phys.\ Rev.\ D {\bf 88}, 024053 (2013).

\bibitem{Brihaye:2012uw}
  Y.~Brihaye and F.~Buisseret,
  Phys.\ Rev.\ D {\bf 87} (2013) 1,  014020.

\bibitem{rajaraman} R. Rajaraman, {\it 
Solitons and Instantons, Volume 15: An Introduction to Solitons and Instantons in Quantum Field Theory}, North-Holland Publishing (1987).

  \bibitem{Gani:2014gxa}
  V.~A.~Gani, A.~E.~Kudryavtsev and M.~A.~Lizunova,
  Phys.\ Rev.\ D {\bf 89} (2014) 12,  125009.
 
  
  \bibitem{vachaspati} T. Vachaspati, {\it Kinks and domain walls}, Cambridge University Press (2006).

  
\bibitem{colsys}
U. Ascher, J. Christiansen and R. D. Russell, Math. Comput. {\bf 33}
(1979), 659; ACM Trans. Math. Softw. {\bf 7} (1981), 209.

 \bibitem{Gani:2015cda}
  V.~A.~Gani, V.~Lensky and M.~A.~Lizunova,
  JHEP {\bf 1508} (2015) 147
 
\bibitem{Copeland:2009as}
  E.~J.~Copeland and M.~I.~Tsumagari,
  Phys.\ Rev.\ D {\bf 80} (2009) 025016
  
  
 \bibitem{Moshir:1981ja}
  M.~Moshir,
  Nucl.\ Phys.\  B {\bf 185} (1981) 318.
 
 
  
\bibitem{Campbell:1983xu}
  D.~K.~Campbell, J.~F.~Schonfeld and C.~A.~Wingate,
  Physica {\bf 9D} (1983) 1.
  
  
  
\bibitem{Campbell:1986mg}
  D.~K.~Campbell and M.~Peyrard,
  Physica {\bf 18D} (1986) 47.
 
  
\bibitem{Campbell:1986nu}
  D.~K.~Campbell, M.~Peyrard and P.~Sodano,
  Physica {\bf 19D} (1986) 165.
  
  

  
\bibitem{Anninos:1991un}
  P.~Anninos, S.~Oliveira and R.~A.~Matzner,
  Phys.\ Rev.\  D {\bf 44} (1991) 1147.
  
  
  

\bibitem{Hoseinmardy:2010zz}
  S.~Hoseinmardy and N.~Riazi,
  Int.\ J.\ Mod.\ Phys.\ A {\bf 25} (2010) 3261.
 


\bibitem{Dorey:2011yw}
  P.~Dorey, K.~Mersh, T.~Romanczukiewicz and Y.~Shnir,
  Phys.\ Rev.\ Lett.\  {\bf 107} (2011) 091602.













  
  
  
  
  
  
   
  
  
  
  
  
  
  
  
\end{thebibliography}
\end{document}